\documentclass{iopart}
\usepackage{iopams,epsfig}

\newtheorem{prop}{Proposition}

\newtheorem{co}{Corollary}
\newtheorem{rem}{Remark}

\newcommand{\sect}[1]{\setcounter{equation}{0}\section{#1}}

\renewcommand{\theequation} {\arabic{section}.\arabic{equation}}

\newcommand{\Sc}{Schr\"odinger}

\newfont{\extra}{msbm10 scaled\magstep1}

\newcommand{\sech}{{\rm sech \,}}
\newcommand{\csch}{{\rm csch \,}}
\newcommand{\sn}{{\rm sn\,}}
\newcommand{\arctanh}{{\rm arctanh\,}}

\def\be{\begin{equation}}
\def\ee{\end{equation}}
\def\bea{\begin{eqnarray}}
\def\eea{\end{eqnarray}}

\def\a{\alpha}
\def\b{\beta}

\def\g{\gamma}

\def\o{\omega}
\def\z{\zeta}

\def\wt{\widetilde}
\def\cs{{\rm csch }\,}
\def\sc{{\rm sech }\,}

\begin{document}

 \large

\title {Second Order Darboux Displacements}

\author{
 B F Samsonov$^{\dag}$\footnote[3]{On leave from
 Physics Department of Tomsk State
 University, 634050 Tomsk, Russia.
 },\
M~L~Glasser$^{\P}$,\ J~Negro$^{\dag}$,
 and  L M Nieto$^{\dag}$ }

\address{\dag\ Departamento de F\'{\i}sica Te\'orica, Universidad de
Valladolid,  47005 Valladolid, Spain}

\address{$\P$\ Department of Physics and Center for Quantum  Device
Technology, Clarkson University, Potsdam NY 13699-5820 (USA)}

\ead{\mailto{samsonov@phys.tsu.ru}, \mailto{laryg@clarkson.edu},
\mailto{jnegro@fta.uva.es},
\mailto{luismi@metodos.fam.cie.uva.es}
 }

\baselineskip=18pt

\begin{abstract}
\baselineskip=16pt
\noindent The potentials for a one dimensional Schr\"odinger
equation that are displaced along the $x$ axis under second order
Darboux transformations, called 2-SUSY invariant, are
characterized in terms of a differential-difference equation. The
solutions of the Schr\"odinger equation with such potentials are
given analytically for any value of the energy. The method is
illustrated by a two-soliton potential. It is proven that a
particular case of the periodic Lam\'e-Ince potential is 2-SUSY
invariant. Both Bloch solutions  of the corresponding \Sc \
equation are found for any value of the energy. A simple analytic
expression for a family of two-gap potentials is derived.
\end{abstract}


\medskip
\medskip

PACS:  03.65.Ge, 03.65.Fd, 03.65.Ca

\medskip
\medskip

\textbf{Corresponding Author}:

L M Nieto

Departamento de F\'\i sica Te\'orica

Universidad de Valladolid

47005 Valladolid, Spain

\medskip

Tel.: 34 983 42 37 54

Fax: 34 983 42 30 13

E-mail: {\it luismi@metodos.fam.cie.uva.es\/}


\newpage

\sect{Introduction}

Originally, it was noticed in the context of periodic potentials
\cite{dunne,fernandez} that transformed Hamiltonians under special
Darboux transformations are displaced with respect to the initial
potential by half a period. Later on, this effect was studied in
more detail  with respect to first order Darboux transformations
(also called 1-SUSY transformations) \cite{mielnik,boris}.
 It was shown that the range of possible displacements produced by a
 Darboux transformation is not restricted to half  a period, but
 can take on values in a set of (in general) complex values called
 ``Darboux displacements" (or SUSY displacements).
 The potentials allowing such displacement are called translationally
invariant
 under the Darboux transformation or, simply, 1-SUSY invariant. It
 was proven that a real-valued even potential function $V_0(x)=V_0(-x)$
 is 1-SUSY invariant if and only if it is of the form
 $V_0(x)=2 \wp (x+\o ')$, where $\wp (x)=\wp(x,g_2,g_3)$
 is the standard Weierstrass elliptic
 function with real and imaginary half-periods $\o $ and $\o '$
 and invariants $g_2$, $g_3$ \cite{Bateman}
 (including its degenerate forms such as $2x^{-2}$,
 $-2\cosh^{-2}x$, and $2\sinh^{-2}x$). This
 result really means that the family of 1-SUSY invariant
 potentials is rather sparse. Moreover,  it was noticed in 
\cite{mielnik} that a
 simple displacement appears to be
 a ``frustrated case" of the Darboux method.
 Nevertheless,  this property has led to
 mathematically nontrivial results and to unexpected link between
  the theory of elliptic functions and
 supersymmetric quantum mechanics \cite{mielnik,boris}.
  From a
 physical point of view a very remarkable property of 1-SUSY
 invariant potentials is that the corresponding Bloch solutions can be
 found analytically for any value of the energy.
 If a  linear
 combination of Bloch solutions is used as the transformation function 
for a simple SUSY transformation,
 it produces an exactly solvable potential
 with locally perturbed periodic structure. We believe these potentials
 could find applications in describing contact effects in
 crystals, or modelling crystals with embedded inclusions.

 In this paper we continue the investigation of SUSY invariant
 potentials, but at the level of second order Darboux transformations.
  The aim of this work is to study what happens if a potential is
  assumed to be 2-SUSY invariant, i.e., a second Darboux
  transformation results only in a displacement
  $V_0(x)\to V_2(x)=V_0(x+d)$. From the very beginning it is
  clear that this condition is  weaker than 1-SUSY
  invariance, since any 1-SUSY invariant potential is obviously
  also 2-SUSY invariant. Actually, as a preliminary result shows
  \cite{boris},
  the family of 2-SUSY invariant potentials is richer than that
   of 1-SUSY invariant ones; at least, it includes a class of 2-soliton
  potentials which are not 1-SUSY invariant. Indeed, as we show
  below, it is even much richer, since a simple Darboux
  transformation over such a potential gives another 2-SUSY
  invariant potential which cannot be reduced to a displaced copy
  of the initial one. Another remarkable property of a 2-SUSY
  invariant potential is that it allows for an analytic
  representation of both linearly independent solutions of the
  corresponding \Sc \ equation at any value of the energy (not
  necessarily from the spectral set).
  Using this property we are  also
  able to generate new exactly solvable potentials with locally
  distorted periodic structure, as  illustrated below.
  Our results lead us to hypothesise that the general
  Lam\'e-Ince potential is invariant under an $n$th order Darboux
  transformation.

 This paper is organized as follows. In the next Section we
 review several properties of Darboux transformations and 1-SUSY
 invariant potentials.
 In Section 3
 we prove necessary and sufficient conditions for a
 potential $V_0(x)$ to be 2-SUSY invariant and point out some
 simple implications of this property. Section 4 has purely
 illustrative character. Here we apply the results of the
 previous section to a well-studied family of 2-soliton
 potentials. Section 5 is devoted to an analysis of the $n=2$
 case of the Lam\'e-Ince potential
 $V_0(x)=n(n+1)\wp (x+\o ')$.
 It should be mentioned that this remarkable potential has attracted much
 attention from mathematicians as well as physicists.
 Without going into further details we refer the interested reader
 to the
 excellent recent papers \cite{Gest} where a vast literature
 is  summarized. We believe that our general results applied to
 the $n=2$ case establish new properties of this remarkable
 potential.
 In particular, we show that it is 2-SUSY invariant and
 give a simple analytic representation of  its linearly independent
 Bloch solutions in terms of the Weierstrass functions.
 Further we give an explicit formula for 1-parameter family of
 two-gap (i.e. having only two finite forbidden and allowed bands)
 potentials.  General properties of such
 potentials have been studied by algebraic-geometrical methods
 (see e.g. \cite{Gest,Bel}). The most striking of our developments is 
that
 we use only elementary means and the well-known properties of
 Weierstrass functions presented in \cite{Bateman}.

\sect{Preliminaries}

Darboux  transformations (also known as SUSY-QM transformations) have 
become an important tool for dealing with spectral problems associated
with the Schr\"odinger equation, especially in one spatial dimension.
Though
 the basic procedure is quite simple, the method has attracted
 increasing
 attention from mathematicians and physicists
 for more than a century
(see e.g.
 \cite{Matveev_Sall,SUSY}).

One begins with a Hamiltonian
 \begin{equation}
 h_0=-\partial^2+V_0(x)\,,\quad \partial\equiv d/dx
 \,, \quad x\in \Bbb R,
 \end{equation}
 (in appropriate units)
whose eigenspace is two dimensional for any eigenvalue $a \in
{\Bbb C} $:
\begin{equation}\label{2}
h_0u(a,x)=a\, u(a,x)\,,\quad h_0\tilde u(a,x)=a\, \tilde u(a,x)\,.
\end{equation}
 As it is well-known, imposition of boundary, summability or
 Bloch type
 constraints selects ``physical" solutions $\psi(E,x)$ and leads to 
physical
 interpretation of the eigenvalue $E$ as a spectral parameter:
\begin{equation}
h_0\,\psi(E,x)=E\, \psi(E,x)\,,\quad E\in \Bbb R  \,.
\end{equation}
 If  $u_1\equiv u(a_1,x)$ is a real and nodeless solution
 of equation (\ref{2}), then the
{\it 1-SUSY partner\/} Hamiltonian
 \begin{equation}\label{1susy}
h_1=-\partial^2+V_1(x),
\qquad
V_1(x)=V_0(x)-2(\ln\; u_1)''
 \end{equation}
 (the prime denotes the derivative with respect to $x$)
is isospectral with $h_0$. (For non-periodic $V_0(x)$ the spectrum
of $h_1$ may change in one point, but we ignore this for the
moment).
 Furthermore, an {\it intertwining} operator $L$ obeying   
$L\, h_0=h_1\, L$
 has the form
 \be\label{1dt}
 L=-\partial+(\log\; u_1)'\,.
 \ee
 Thus, the eigenfunctions of
 $h_1$ for eigenvalue $a\ne a_1$ are $v(a,x)=L\,u(a,x)$,
 $h_1v(a,x)=av(a,x)$.
 This procedure can
 be repeated by starting with $h_1$ to produce a second isospectral 
Hamiltonian
$h_{2}=-\partial^2+V_{2}(x)$, etc. It can be shown that in this
way, for fixed $u_1$  and $u_2\equiv u(a_2,x)$, the transformed
potential is expressed in terms of the
 Wronskian $W(x)\equiv W(u_1,u_2)=u_1u_2'-u_1'u_2$ as
 \begin{equation} \label{v12}
 V_{2}(x)=V_0(x)-2[\log\; W(x)]'' \ .
 \end{equation}
For $a\ne a_1,a_2$ the
  eigenfunctions of $h_{2}$ are
 $v(a,x)=
 W(u_1,u_2,u )\,W^{-1}(u_1,u_2)$,
 $h_2v(a,x)=av(a,x)$
 where $u\equiv u(a,x)$.
 It is easy
 to see that $W(u_1,u_2,\tilde u_{1,2})\propto u_{2,1}$
 so that for $a= a_1,a_2$, up to an inessential constant factor, this 
formula gives
 \be\label{particular}
 v_{1,2} \equiv v_{1,2}(a_{1,2},x)
 = u_{2,1}W^{-1}(u_1,u_2)\,,\quad
 h_2v_{1,2}=a_{1,2}v_{1,2} \,.
 \ee
 We note that this is a particular case of a much more general result for
 a chain of $N$ Darboux transformations
\cite{SUSY}.
 Solutions of the initial \Sc \ equation $u_j\equiv u_j(a_j,x)$ are 
called
 {\it transformation functions} while their eigenvalues $a_j$ are known 
as {\it factorization constants}. We refer the reader to \cite{SUSY}
 for more details.

A particularly interesting case is where $V_0$ is periodic:
\begin{equation}
V_0(x+T)=V_0(x) \, .
\end{equation}
We recall that inside the eigenfunction space of each eigenvalue
for such a Hamiltonian we can build up  Bloch functions, i.e., two
 linearly independent solutions $u^{\pm}$ of $h_0\,u^\pm = a\,u^\pm $
 such that
 \begin{equation}\label{betas}
 u^{\pm}(a,x+T)=[\beta(a)]^{\pm 1}\,u^{\pm}(a,x)\,,\quad \beta (a)\in 
\Bbb C
 \,.
 \end{equation}
 When for $a=E$ one has $|\beta(E)|=1$, the $E$--values
 belong to a spectral band
 and the corresponding
 (``physical") solutions are bounded, while for
 $|\beta(a)|\ne 1$ they are unbounded (``non-physical")
 and the values of $a$ lay in forbidden
  bands (see e.g. \cite{magnus}).
   By using appropriate Bloch functions as
  transformation functions in the Darboux algorithm
 one can construct periodic partner Hamiltonians
 \cite{dunne}--\cite{boris}, \cite{Trl}. A linear combination of Bloch
 functions leads to a perturbed  periodic structure which
 is asymptotically periodic.

 Let, for instance, the transformation functions be
Bloch functions
 $u_1^+(x)$ and  $u_2^-(x)$
 with corresponding factors $\b_{1,2}=\b (a_{1,2})$ as defined in
 (\ref{betas}).
 Then, according to (\ref{particular}), Bloch
 solutions of the transformed equation corresponding to
 eigenvalues $a_1$ and $a_2$ are respectively
 \be \label{v12f}
 v_1^-=\frac{u_2^-}{W(x)}\,,\quad v_2^+=\frac{u_1^+}{W(x)}\,, \quad
 h_2v_{1,2}^{\pm}=a_{1,2}v_{1,2}^{\pm}\, ,
 \ee
 where
 \be\label{BlochV}
v_{1}^-(x+T)={\beta^{-1}_{1}}v_{1}^-(x)\,,\quad v_{2}^+(x+T)=
 \beta _{2}v_{2}^+(x) \,.
 \ee

 If under a Darboux transformation we have
\begin{equation} V_1(x)=V_0(x+d)\end{equation}
then $d$ is called a {\it Darboux displacement}
 and $V_0$ is said to be {\it 1-SUSY invariant}.
 The first cases of
Darboux displacements were found for periodic potentials
\cite{dunne,fernandez}.
 Later it was proven \cite{mielnik} that an even function
$V_0(x)$ allows for a first order Darboux displacement if and only
if it satisfies the nonlinear differential-difference equation
\begin{equation} \label{11}
V_0(x)+V_0(x+d)-\frac{1}{2}
\left[\frac{V_0'(x)+V_0'(x+d)}{V_0(x)-V_0(x+d)}\right]^2=
\mbox {const}\, ,
\end{equation}
which, up to a constant, is equivalent to the addition formula for
the Weierstrass $\wp$ function. This result means that the family
of even 1-SUSY invariant potentials is restricted  to the  $\wp$
function and its degenerate forms, such as one-soliton potentials.
 In the next Section we prove  necessary and sufficient
 conditions for a potential $V_0$ to admit a displacement under a
 second order Darboux transformation. The potentials possessing
 this property will be called {\it  2-SUSY invariant}.

\sect{Second order Darboux displacements}

\vskip .1in

Let, as before, $u_1^+(x)$ and $u_2^-(x)$ be two Bloch
 eigenfunctions of $h_0$
 which are chosen as transformation functions for a
 2-SUSY transformation.
 Although we concentrate  on the case of  periodic potentials
 $V_0(x+T)=V_0(x)$, our results have a more general
 character, as will be mentioned below.
  Now we assume that the potential $V_{2}$,
 obtained according to (\ref{v12}), is displaced from the original one, 
i.e.,
 \begin{equation}\label{v2d}
 V_{2}(x)=V_0(x+d)\,.
 \end{equation}
Thus, the differential equations corresponding to $h_0$ and $h_{2}$
have exactly the same solutions, but shifted by the displacement
 $d$.
 Hence, taking into account the property (\ref{BlochV}) we see that
 \be\label{uv}
 v_1^-(x-d)\propto u_1^-(x)\,,\quad
 v_2^+(x-d)\propto u_2^+(x) \, .
 \ee
>From this and (\ref{v12f}) it follows readily that
 \begin{equation} \label{16}
 W(x)=c_1\frac{u_2^-(x)}{u_1^-(x+d)}
 =c_2\frac{ u_1^+(x)}{u_2^+(x+d)}\, ,
 \end{equation}
 where $c_{1,2}$ are constants.
It is convenient to use the notation
  \begin{equation} \label{17}
 \wt{x}=x+d\,,
 \mbox{\hskip .1in}
 \Phi(x)=1/W(x),
 \mbox{\hskip  .1in}
 p\,(x)=-2[\ln\;\Phi(x)]'\,.
 \end{equation}
 Therefore,
 $p^\prime (x)=V_0(x)-V_0(\wt{x})$ and from (\ref{16}) it follows
 that
 \begin{equation} \label{tilde}
u_1^-(\wt x)=c\,\Phi(x)\,u_2^-(x) \, ,
 \end{equation}
where $c$ is an inessential non-zero constant. After taking the
second derivative of (\ref{tilde}) and using (\ref{2}), one
obtains
 \begin{equation}\label{u2}
 [V_0(\wt{x})-a_1]\Phi(x)u_2^-(x)=
 [V_0(x)-a_2]\Phi(x)u_2^-(x)+u_2^-(x)\Phi''(x)+2{[u_2^-(x)\,]}'\Phi'(x)
\ \
 \end{equation}
 Now, from (\ref{17}) one has
\begin{equation} \Phi'=-\frac{p}{2}\,\Phi\,,\mbox{\hskip .3in} \Phi''
=-\frac{p'}{2}\,\Phi+\frac{p^2}{4}\,\Phi\, ,
 \end{equation}
 and from (\ref{u2}) this yields
 \begin{equation}\label{21}
 [\ln u_2^-(x)]'=\frac{p'}{2p}+\frac{p}{4}+
 \frac{a_1-a_2}{p}\, .
 \end{equation}
 Similarly
 \begin{equation}\label{22}
 [\ln u_1^+(x)]'=\frac{p'}{2p}+\frac{p}{4}-
 \frac{a_1-a_2}{p}\, ,
 \end{equation}
 \begin{equation}
 [\ln u_2^+(\wt{x}\,)]'=\frac{p'}{2p}-
 \frac{p}{4}-\frac{a_1-a_2}{p}\, ,
 \end{equation}
 \begin{equation} \label{22b}
 [\ln u_1^-(\wt{x}\,)]'=\frac{p'}{2p}-
 \frac{p}{4}+\frac{a_1-a_2}{p}\,.
 \end{equation}
\begin{rem}
The formulas $($\ref{21}$)$--$($\ref{22b}$)$ illustrate one of the
most remarkable properties of a 2-SUSY invariant potential. The
four Bloch functions corresponding to the eigenvalues $a_1$ and
$a_2$ producing a displacement are expressible only in terms of
the potential difference $p'(x)=V_0(x)-V_0(x+d)$ $($a known
function\,$)$ and its primitive $($also known from Eq.\
$($\ref{29}\,$)$ below\,$)$.
\end{rem}
Later on for a
particular case of the Weierstrass potential we will integrate
these equations to get analytic expressions for the solutions
of the \Sc \ equation.

We turn next to finding necessary and sufficient conditions
for $V_0(x)$ to admit a displacement under
a second order Darboux transformation
(shortly second order Darboux displacement).

\noindent
\begin{prop}
The Hamiltonian $h_0$ allows for a second order Darboux
displacement if and only if the potential $V_0(x)$ satisfies the
following differential--difference equation$:$
 \be\label{th}
 T -  \frac{1}{4} \left[\frac{ T '}{M}\right]^2= \rm{constant}\, ,
 \ee
 where
 \be   \label{T}
 T=(S'/D)^2+4S-2D''/D, \qquad M= D-(S'/D)'\, ,
 \ee
 and
\be\label{q}
S=V_0(x)+V_0(\wt x),\qquad D=V_0(x)-V_0(\wt x)\,.
\ee
\end{prop}
Note that this formula
resembles (2.13) for the first order displacement, but the
functions $T$ and $M$ involved depend in a much more complex way
on $V_0(x)$.

\noindent
{\it Proof}.
 Since
$u''=[(u'/u)'+(u'/u)^2]u$, the Schr\"odinger equation for $u_2^-$
is
\begin{equation} \label{25}
\left[\frac{{(u_2^-)}'}{u_2}\right]'+
\left[\frac{{(u_2^-)}'}{u_2}\right]^2+a_2-V_0(x)=0 \, ,
 \end{equation}
 which together with (\ref{21}) results in
\begin{equation}\label{26}
2pp''-(p')^2+\frac{1}{4}p^4+4(a_1-a_2)^2+2p^2(a_1+a_2-S) =0 \, ,
 \end{equation}
where $S$ is defined in (\ref{q}).
  Now we take the
 derivative of (\ref{26}) and, since $p'=V_0(x)-V_0(x+d)$ is a
 known function, we can treat the  result as a quadratic equation
 for $p$, whose solution is
 \begin{equation} \label{27}
 p=\frac{S'}{p'} \pm \sqrt{\left(\frac{S'}{p'}\right)^2
 -2\frac{p'''}{p'}+4S-4(a_1+a_2)}\,.
 \end{equation}
 Then taking the derivative of
(\ref{27}), and rearranging, we obtain the
 nonlinear differential-difference  equation for $V_0(x)$
 \begin{equation} \label{28}
 4(a_1+a_2)=T-\frac{1}{4}
 \left[\frac{T'}{D-(S'/D)'}\right]^2
 \end{equation}
 where $T$ is defined by (\ref{T}).
 Keeping in mind that the eigenvalues $a_1$ and $a_2$ are
 independent of $x$ and taking into consideration the definition
 of  $M$ (second formula in (\ref{T})), we see that the necessary
 condition is fulfilled. We shall show now
 that this condition is also sufficient.

 First, we notice
 that by using (\ref{28}) to eliminate $a_1+a_2$ in
(\ref{27}) we find
\begin{equation} \label{29}
p=\frac{S'}{D}+
\frac{1}{2}\frac{T'}{D-(S'/D)'}\, .
 \end{equation}
Here we have to choose only plus sign in (\ref{27}) since
 it is easy to
see that the minus contradicts the condition
$p'=V_0(x)-V_0(x+d)$.
 If the right hand side of
 (\ref{28}) is constant, then from (\ref{29}) we obtain $p$ which, by
 retracing our steps, satisfies (\ref{26}) which is equivalent to the
 \Sc \ equation (\ref{25}) for
$u_2^-$. Therefore, from (\ref{21}) and (\ref{22}), we have the
logarithmic derivatives of
$u_2^-$ and $u_1^+$. These being known, we are able to calculate
 the second logarithmic derivative of their Wronskian
 \begin{equation}
 [\,\ln\; W\,]''=(a_1-a_2)
 \frac{[\,\ln (u_1^+u_2^-)\,]'}{[\,\ln ( u_1^+/u_2^-)\,]'}-
 \left(\frac{a_1-a_2}{
 [\,\ln(u_1^+/u_2^-)\,]'}\right)^2=\frac{1}{2}D\, ,
 \end{equation}
 which by (\ref{v12}) means that $h_0$ possesses a second order
 Darboux displacement.            \hfill$\Box$

\begin{co}
Any 2-SUSY invariant potential allowing a set of displacements $d$
generates a one-parameter family of 2-SUSY invariant potentials
$($some of them may have singular points$)$ obtained by a
 1-SUSY transformation with one of the four function $u_1^\pm$, $u_2^\pm$
 as transformation function.
\end{co}

 \noindent {\it Proof}.
Let $V_0(x)$ be 2-SUSY invariant. This means that there
 exists a second order Darboux transformation with transformation
 functions $u_1^-$ and $u_2^+$ which results only
 in a displacement of the transformed potential
 $V_2(x)=V_0(x+d)$.
 Any second order transformation may be factored into
 two successive first order transformations. One
 can realize the first transformation with the transformation
 function $u_1^-$ to get an intermediate potential $V_1(x)$ which
 cannot be obtained by a displacement of $V_0(x)$ if the later is not
 1-SUSY invariant and hence $V_1(x)$ is essentially different of
 $V_0(x)$. We shall show that
$V_1(x)$ allows for the same second order Darboux displacement.
In order to
 show this property take the logarithms of the first equality in
  (\ref{16})
 \[
 2\log W(x) + 2 \log u_1^-(x+d) = 2\log c_1 + 2 \log u_2^-(x)  \,.
 \]
 After taking two derivatives and using (\ref{v12}) and (\ref{v2d}) we
 obtain
 \be\label{family}
V_0(x) - 2 [\log u_2^-(x)]''= V_0(x+d)-2 [\log u_1^-(x+d)]'' \,.
 \ee
The left hand side of this equality means that we realize a 1-SUSY
transformation of the potential $V_0$ with the transformation
function $u_2^-(x)$ and from the right hand side we learn that a
similar transformation is realized of the shifted version of the
same potential with the transformation function $u_1^-(x+d)$.
Since the Darboux transformations are invertible, we see that
these transformed potentials may always be related to each other
with the help of a 2-SUSY transformation for which $V_0$ is an
intermediate potential. Hence, the equality (\ref{family}) means
that after such a transformation we get only a displacement by the
value $d$.
 If the function $u_2^-(x)$ has nodes, the 1-SUSY partner of $V_0(x)$
 will have singularities (poles). To displace $V_0(x)$ by another
 value we have to choose a second pair of transformation functions
 $u_1^+$ and $u_2^-$ that give us another member of the
 family of 2-SUSY invariant potentials.
 \hfill $\Box$

 \begin{rem}
 No member of the family obtained above is a
 displaced copy of the initial potential if it is not 1-SUSY
 invariant. Moreover, few of the potentials are displaced
 copies of other members of the same family,
 which means that the whole set of 2-SUSY invariant potentials is
 quite rich.
 \end{rem}
 Indeed, if the initial potential is not 1-SUSY invariant it is
 not possible to displace it by a 1-SUSY transformation.
 Next, any
 two different members of this family are related by a 2-SUSY
 transformation, where $V_0$ is an intermediate potential.
 The factorization parameters of such a transformation are independent
 of each other, which
 means that once a representative of the family is fixed, $\wt V$, the
 whole family is characterized by two independent (factorization)
 parameters. The potential  $\wt V$, being 2-SUSY invariant, also admits
 2-SUSY transformations resulting only in a displacement, but
 these transformations are characterized by one parameter only.
 This follows from equations $($\ref{26}$)$ and $($\ref{28}$)$ which may
 be considered as defining one of the factorization constants $($say
 $a_2$$)$ as an implicit function of the other $(a_1)$.
 This
 property will also be illustrated in the next Section and used in
 the Section 5 to produce new exactly solvable periodic
 potentials.

\begin{rem}
The results of this Section are valid not only for a periodic
potential but for any potential satisfying the conditions of
Proposition~1.
\end{rem}
This follows from the fact that the only place
where we used the Bloch
 property of solutions of the \Sc \ equation
 (that is the periodicity of the potential $V_0(x)$)
 is in formula
$($\ref{uv}\,$)$. Thus, $u_1^\pm$ and $u_2^\pm$ can be not only
Bloch solutions, but two pairs of linearly independent solutions
of the \Sc \ equation with a non-periodic potential.

A simple sufficient condition for (\ref{th}) is that
$T=(S'/D)^2+4S-2D''/D$ be constant. For example,  consider the
1-SUSY invariant potential
 $V_0(x)=2\wp(x-\o' )$.
 From the differential equation for $\wp$, we
obtain $V_0''=3V_0^2-g_2$ and therefore
$D''=3V_0''(x)-3V_0''(x+d)=3DS$. From (\ref{11}), the 1-SUSY
invariance requires $(S'/D)^2=2S+\mbox{const}$. Therefore, in this
case, $T=(S'/D)^2-2S=\mbox{const}$
 and, consequently, $V_0$ is
also 2-SUSY invariant, which of course was evident from the
outset.

Note that for even potentials, $V_0(-x)=V_0(x)$, the expression in
square brackets in (\ref{th}) is not defined for $x=-d/2$. Using
Mathematica \cite{Mathematica} we have found that in this case
$$
\lim_{x\to d/2}\frac{1}{4}
 \left[\frac{T'}{D-(S'/D)'}\right]^2 =
 \frac{12(V_0')^3V_0''+(V_0'')^2V_0'''-V_0'
 [(V_0''')^2+V_0''V_0^{(\mbox{iv})}]+(V_0')^2V_0^{(\mbox{v})}}{V_0'[6(V_0')^3+
 V_0''V_0'''-V_0'V_0^{(\mbox{iv})}]} \, ,
$$
where all the functions in the right hand side are evaluated at $x=d/2$.

\sect{\bf Two-soliton potentials}

In this section we illustrate our previous developments for a well-studied
two-soliton potential.
In its most general form this potential is determined by four
 parameters:
 \be \label{2soliton}
 V_0(x)=\frac{2(\a_1^2-\a_2^2)
 (\a_1^2\,\sech ^2(\a_1x+\b_1)+\a_2^2\,\csch ^2(\a_2x+\b_2)}%
 {[\a_1\tanh (\a_1x+\b_1)-\a_2\coth (\a_2x+\b_2)]^2}\, .
 \ee
 The  parameters $\a_2>\a_1>0$ define the positions of the
 discrete levels  $E_{0,1}=-\a_{2,1}^2$, while $\b_1$ and $\b_2$
 characterize isospectral deformations.
 It has been shown by  direct calculation \cite{boris}
 that the symmetrical case, $\b_1=\b_2=0$, is 2-SUSY invariant.
 We note that exactly the same calculations can be carried out
 for the general case, indicating that for any fixed values of the
 parameters $\a_{1,2}$ and $\b_{1,2}$ the potential
 (\ref{2soliton}) is 2-SUSY invariant.
  This illustrates Remark 3.
 With  evident
 modifications of the results of \cite{boris} we give the
 solutions $u_1^+$ and $u_2^-$ producing the second order Darboux
 displacement $d$ of $V_0$:
 \[
 u_{1,2}^\pm=W(u_{10},u_{20},u_{30,40})W^{-1}(u_{10},u_{20}),
 \]
where
 $u_{10}=\cosh (\a_1x+\b_1)$, $u_{20}=\sinh(\a_2x+\b_2)$ are
 solutions of the free particle \Sc \ equation generating the
potential (\ref{2soliton}) from the zero potential, and
 $u_{30,40}=\exp (\a_{3,4}x)$, $\a_{3,4}>0$.
 When parameters $\a_3$ and $\a_4$  vary in a certain domain,
  independently of each other,
 these functions produce an isospectral deformation of
  $V_0$, resulting in the same expression with the different
 values of $\b_1$ and $\b_2$. To select from these transformations
 only ones leading to the shift $V_0(x)\to V_2(x)=V_0(x+d)$
 one has to impose an additional restriction on $\a_3$ and $\a_4$.
 The analysis in  reference \cite{boris} shows that
 to get real displacements one has to
  change
  the factorization constants $\a_3$ and $\a_4$
  so that they fall
   between the existing discrete levels
  of $V_0$:
    $\a_1<\a_3<\a_4<\a_2$.

Let us now characterize the one-parameter family mentioned in
Corollary 1. For this purpose we have to realize a 1-SUSY
transformation of the potential $V_0$ by using one of
$u_{1,2}^\pm$ as the transformation function.
 For simplicity let us fix $\b_1=\b_2=0$ and choose the function
 $u_1^+$ to produce the potential $V_1$.
 According to properties of Darboux transformations (see e.g.
 \cite{SUSY}) this is equivalent to a third order
 transformation of the zero potential with transformation
 functions $u_{10}$,  $u_{20}$ and  $u_{30}$ or simply to a chain of
 transformations where, for instance, the first transformation is
 realized with the function
 $u_{30}$.
For such a transformation the zero
 potential is not affected, but the functions $u_{10}$ and $u_{20}$
 receive changes and, up to a constant factor, become
  $u_{10}(x)\to v_{10}(x)=\cosh (\a_1x+\g_1)$,
 $u_{20}(x)\to v_{20}(x)=\sinh (\a_2x+\g_2)$, where
\be\label{g12}
 \g_{1}=\arctanh(\a_{1}/\a_3)\,,\quad
  \g_{2}=\arctanh(\a_{2}/\a_3)\,.
 \ee
 After the second order transformation with the functions
 $v_{10}$ and  $v_{20}$ we get from the zero potential the same
 two-soliton potential (\ref{2soliton}) where $\b$'s are replaced
 by $\g$'s.
 This is just the 1-parameter family of Corollary 1 where $\a_3$
 is the parameter.
 From the definition (\ref{g12}) we
 see that the parameters $\g_1$ and $\g_2$ are not independent of
 each other: $\tanh\g_1/\tanh\g_2=\a_1/\a_2$. They  vary so
 that the potential $V_1$ cannot be reduced to a displaced copy
 of the potential (\ref{2soliton}) at $\b_1=\b_2=0$, since the later
 condition requires $\a_1/\a_2=\g_1/\g_2$, which leads to
  $\tanh\g_1/\tanh\g_2=\g_1/\g_2$. The later is possible only
 for $\g_1=\g_2$ which contradicts to $\a_1\ne \a_2$.
 Just from this, the potential $V_1$, being of course
 a two-soliton potential, is essentially different from $V_0$, as
 pointed out in Remark 2.

We will illustrate other properties derived in the previous
section for the simplest case of this potential. We choose
$\a_1=1$ and $\a_2=2$ to produce the well-known P\"oschl-Teller
potential well
 \be\label{PT}
 V_0(x)=-6 \,{\rm sech}^2x\,
 \ee
 for which $E_0=-4$ and $E_1=-1$.

 From (\ref{T}) we find  the value of $T$,
 \be
T=    4\,\left[ -1 + {\mbox{csch}^2d} - 4\,{\mbox{sech}^2x} -
     4\,{\mbox{sech}^2(d + x)} \right]\, ,
\ee
 which after being substituted into Eq. (\ref{28}) gives the
 $x$-independent quantity
 \be\label{suma}
 a_1+a_2=-5-3\,\mbox{csch}^2d\, .
 \ee
 This means that the necessary and
 sufficient conditions for $V_0$ (\ref{PT}) to be 2-SUSY invariant are
 fulfilled.
 Now from (\ref{29}) we obtain $p(x)=-6[\tanh x+\csch d\, \cosh
x\,\sech (x+d)]$.
 From Eq. (\ref{26}) we deduce the difference of the factorization
 constants
\be\label{dif}
a_1-a_2=\pm (3\coth d\,\sqrt{1-3\csch^2 d\vphantom {I^I}}\,).
\ee
 From here and (\ref{suma}) one gets one factorization constant as
 a function of the other:
 $a_1=\frac 12(-6-a_2\pm  \sqrt{-12a_2-3a_2^2\vphantom{I^I}}\,)$.
 It is clear that this equation and (\ref{suma})
 define the displacement $d$ as a function of $a_2$:
 \be
 d=\mbox{arccsch}\frac{\sqrt{-4-a_2+\sqrt 3\sqrt{-a_2(a_2+4)}}}
 {\sqrt  6}\,.
 \ee
 It follows from here that real displacements are possible
 only when both factorization constants lie between discrete
 levels, $-4<a_{1,2}<-1$, and in the above formula for $a_1$ the
 lower sign is chosen. This agrees completely
 with the previous result \cite{boris}.
 We see also that when $a_1\to -3$,
 then $a_2\to -3$,
and we get the minimal value of the displacement $d_{\rm min}= {\rm
arccsh\,}(1/\sqrt{3})$. The other limit is $a_1\to -4$, which gives
$a_2\to -1$, so that the displacement increases indefinitely, $d\to
\infty$. In this respect we remark that for 1-soliton potentials the (first
order Darboux) displacements range is always $(0,\infty)$.

 We are able now to find the right hand side of equation
 (\ref{21}). It is expressed in terms of elementary functions only
 and its primitive gives us a solution of the \Sc \ equation with
 the potential $V_0$:
$$
  u_2^-(x)=\cosh (x+d)
  e^{-\frac 12(3\coth d+
  \sqrt{1-3{\rm csch}^2d }\,)}
  [3-\cosh 2d+\cosh 2x+\cosh (2x+2d)]^{1/2}
$$
\be    \label{u2pl}
 \times\, {\rm sech}^2 x
  \left[
  \displaystyle{\frac%
  {\sqrt 2(\coth d+2\tanh x)\sinh d+\sqrt{\cosh 2d-7}}
  {-\sqrt 2(\coth d+2\tanh x)\sinh d+\sqrt{\cosh 2d-7}}}
  \right]^{1/2} \, .
 \ee
 The derivative of this expression yields the potential
 difference as a function of the parameter $d$:
$$
 \Delta V=
 [(4\cosh x\,\cs d\, \sc (x+d)+\tanh x )
 (\sc^2(x+d)\tanh (x+d)+2\sc^2x \tanh  x)
 $$
 \bea  \label{deltaV}
+\,  (\sc^2(x+d)-\sc^2x)(3\cs^2d+\coth d\,\sqrt{1-3\cs^2d}\\[.5em]
 - \, \sc^2x-2\sc^2(x+d)+3\tanh^2x)]   [\tanh x+\cs d\cosh x\, \sc 
(x+d)]^{-2} \,.
\nonumber
\eea
 Of course, this intermediate quantity can be eliminated from the final
 expression to obtain a somewhat simpler formula for the family of
 two-soliton potentials:
 \bea
 V_1(x)&=&12[4+9a_2-a_2^2-6\,\sc^2x+12\sqrt{a_2}\tanh x   \nonumber \\
&& -(a_2-1)((a_2+4)\cosh 2x +4\sqrt{a_2}\sinh 2x)]  \nonumber   \\
 & &\   \times\,
  [a_2-4+(a_2+2)\cosh 2x+3a_2\sinh 2x]^{-2} \, .
 \eea

\begin{center}
\epsfig{file=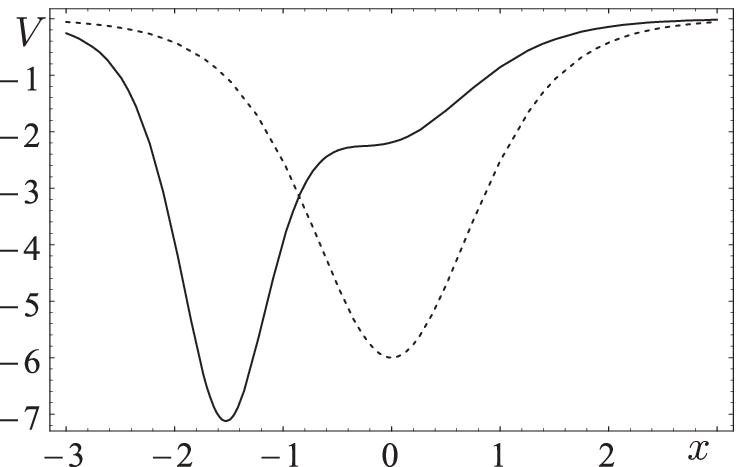}  
\medskip

\begin{minipage}{14cm}
{\small Figure 1:
 Two soliton P\"oschl-Teller potential (dashed line)
  and its SUSY partner at $a_2=-4.1$.
}
\end{minipage}
\end{center}
\medskip

 For the potential difference to be regular one has to choose
  the factorization constant $a_2$ to be less than the ground state level of
  $V_0$, $a_2<-4$, which gives complex values for the parameter $d$.
  Despite this, the potential difference is real and the solution
  (\ref{u2pl}) is also real (and nodeless) if it is set real at any point, 
i.e. $u_2^-(0)=1$.
  As an example we have plotted one of the potentials
 $V_1=V_0+\Delta V$ with $\Delta V$ as in
  (\ref{deltaV})
  in the Fig.\ 1 (solid line) along with $V_0$ (dashed line).

\sect{The Lam\'e-Ince potential}

A much more interesting example of second order Darboux
 displacements is given by the
 following result.
\begin{prop}
The periodic two-gap potential
 \be \label{wp}
 V_0(x)=6\wp(x+\o ')
 \ee
 where $\o '$ and $\o $ are the imaginary and real half-periods of
 the Weierstrass function $\wp$, with $g_2$ and $g_3$ as
 invariants,
allows for second order Darboux displacements. The eigenvalues $a_{1,2}$
of the transformation functions producing the displacement $d$ are
given by
 \be\label{34}
 a_{1,2}=-\frac{3}{2}\,[\,\wp(d)\mp\sqrt{g_2-3\wp^2(d)\vphantom{i^i}}\,]\,.
 \ee
 \end{prop}

\noindent
{\it Proof}.
 The proof consists simply in calculating the left hand side of
 the equation (\ref{th}). We obtain first some useful relations
 between our variables which are direct implications of the
 well-known properties of the Weierstrass functions (see e.g.
 \cite{Bateman}).
 From the differential equation for $\wp$ one gets $D''=SD$, so
$T=(S'/D)^2+2S$. The addition formula for $\wp$ is
\begin{equation} \label{addition}
\wp(u+v)+\wp(u)+\wp(v)=\frac{1}{4}
\left[\frac{\wp'(u)-\wp'(v)}{\wp(u)-\wp(v)}\right]^2\,.
\end{equation}
By setting $u=x+\o '+d$, $v=-x-\o '$ and noting that $\wp$ is
even, this translates into $(S'/D)^2=\frac{2}{3}S+4\wp(d)$ and
$T=\frac{8}{3}S+4\wp(d)$. By taking the square root of the former
of these relations and differentiating, we get
\begin{equation}
\frac{1}{D}\left(\frac{S'}{D}\right)'=\frac{1}{3}\end{equation}
and $T'=8S'/3$, $D-(S'/D)'=2D/3$. Inserting these relations into
(\ref{27}) yields
 \be \label{p-function}
 p=3(S'/D).
 \ee
 Now
 we are able to calculate the left hand side of (\ref{th}) which
 gives
 \begin{equation}\label{theorem}
 a_1+a_2=-3\wp(d)\,.
 \end{equation}
 This proves the first part of the statement. Now we shall find
 the eigenvalues of the transformation functions. For this purpose
 we shall use the equation (\ref{26}).
Since $[\wp'(x)]^2=4\wp^3(x)-g_2\wp(x)$, we have the
 algebraic identity
 \begin{equation}\label{identity}
 D^2-6\frac{S'D'}{D}+3S^2=36g_2 \, ,
 \end{equation}
 and since $p=3S'/D$, inserting it and (\ref{theorem})
 in (\ref{26}) and taking into account (\ref{identity}) we obtain
\begin{equation}\label{dif2}
(a_1-a_2)^2=9[g_2-3\wp^2(d)]\,.
\end{equation}
 This equality together with (\ref{theorem}) proves the statement.
One can check that formulas (\ref{theorem}) and (\ref{dif2}), for
the sum and difference of factorization constants, reduce to those
for the specific case of the two-soliton potential (\ref{suma})
and (\ref{dif}), respectively, when we take the half periods
$\o'=i\pi/2$ and $\o =\infty$.
 \hfill $\Box$

Some straightforward consequences of Proposition 2 are:

\noindent {\bf Corollary 2} {\em Three of the five band edges for
the potential $($\ref{wp}$)$},
\be E_1=3e_3\,,\quad
E_{1'}=3e_2\,,\quad E_2=3e_1 \ee {\em correspond to  $d=\omega $,
$d=\omega +\omega '$ and $d=\omega '$, where} \be e_1=\wp (\omega
),\quad  e_2=\wp (\omega +\omega '), \quad e_3=\wp (\omega ')\, .
\ee
{\em  The lowest $E_0=-\sqrt{3g_2\vphantom{i^i}}$ and the
highest
 $E_{2'}=\sqrt{3g_2\vphantom{i^i}}$
 band edges are stationary points for the factorization constants
 $a_{1}$ and $a_2$ as
 functions of the displacement $d$.}

\noindent {\bf Corollary 3}
{\em The pair of $($real\,$)$ factorization constants which
give rise to a real displacement by means of a second order
 Darboux transformation belong to the first finite forbidden band
$ [E_{1},E_{1'}]$. The real displacements so produced are in the
interval $(d_{{\rm min}},d_{{\rm max}})$, where
 \be
 d_{{\rm  min}}=\wp^{-1}\big(\sqrt{g_2/3}\, \big)\,, \qquad
 d_{{\rm max}}=\omega.
 \ee
 $\wp ^{-1}(t)$ is the function inverse to $t=\wp (z)$
 for which an appropriate sheet has to be chosen;
 $d_{{\rm max}}=\omega $ is realized for $a_2=E_1$ and
 $a_1=E_{1'}$, while $d_{{\rm min}}$ is realized for
$a_1=a_2=-\sqrt{3\,g_2}/2$. }

\noindent It is interesting to note that $-e_k$ are the positions
of the band edges for
the one-gap potential $V_0=-2\wp (x+\omega ')$ and that in this case the 
range of the (first order) Darboux displacements is $(0,\infty)$.

\noindent
{\it Proof}. The first part  of  Corollary 2 follows immediately
from Proposition 2 and known properties of the quantities $e_k$
given in Ref.\ \cite{Bateman} (vol.\ 2), when one uses $d=\omega $,
$d=\omega +\omega '$ and $d=\omega '$. The second one is a direct
consequence of (\ref{34}) and conditions $a'_{1,2}(d)=0$. Corollary 3
follows from the restriction on $d$ to be such that
$g_2-3\wp ^2(d)$ be positive, since according to  Proposition 2 it
is inside of the square root. \hfill $\Box$

\stepcounter{co} \stepcounter{co}

 It is not difficult to see that for $d=\omega$,
 the factorization constants are $a_2=E_1$ and $a_1=E_{1'}$.
 This is the
 only possibility for the transformed potential to be displaced by
 half a real period of the Weierstrass $\wp$ function.
 The possibility for the potential (\ref{wp}) to be displaced by a
 half-period was previously noticed in
 \cite{dunne,fernandez}.
 Here we indicate a possibility
 of displacing the argument of the potential (\ref{wp}) by
 a complex value and, in particular, by half an imaginary period.
 We remark  that the values for the band edges are given in
 \cite{fernandez} for the Jacobi form of the Schr\"odinger equation,
 which
 we relate to the Weierstrass form
 in Appendix A.

We are now able to integrate equations (\ref{21})--(\ref{22b})
and find analytic expressions for the Bloch functions, as it is established 
in the following proposition.

\begin{prop}
The functions
 \bea
 &&u_2^-(x)=\pm \sqrt p \left( \frac{\sigma
 (x+\omega '+d)}{\sigma (x+\omega')} \right)^{3/2} \left(
 \frac{\sigma (x-x_{1})}{\sigma (x-x_{2})} \right)^{1/2}
 e^{\,(\,b\,-\,\frac 32\,\zeta (\,d\,))\,x} \label{45}\\
 && u_2^+(\wt x )=\frac{p\,(x)}{u_2^-(x)}\label{46}
 \eea
are the Bloch eigenfunctions for the Hamiltonian $($\ref{wp}$)$ with
 the eigenvalue $a_2$, and
\bea\label{47}
 &&u_1^-(\wt x )=u_2^-(x)
\left( \frac{\sigma (x+\omega ')}{\sigma (x+\omega'+d)}
\right)^{3}
e^{\,3\,\zeta (\,d\,)\,x} \\
&&u_1^+(x)=\frac{p\,(x)}{u_1^-(\wt x )}\label{48}
\eea
 are the
Bloch eigenfunctions with the eigenvalue $a_1$.
Here $\sigma $ and $\zeta $ are standard Weierstrass functions,
 \be  \label{49}
 \begin{array}{l}
 x_1=\wp^{-1}\left( -\frac 12\wp_0+
 \frac 12\sqrt{g_2-3\wp_0^2} \right)-\omega '\\
 x_2=\wp^{-1}\left(
 -\frac 12\wp_0-\frac 12\sqrt{g_2-3\wp_0^2} \right)-\omega '
 \end{array}
 \ee
 $\wp_0=\wp(d)$, $\wp^{-1}$ is the function inverse to $\wp $ and
 \be\label{50}
 b=\frac 12\,\zeta (\omega '-x_2)
 -\frac 12\,\zeta (\omega  '-x_1)\,.
 \ee
 \end{prop}
 The proof can be found in Appendix B.

\begin{rem}
The solutions given in $($\ref{45}\,$)$ and  $($\ref{46}\,$)$ are
valid for any eigenvalue $a_2$. By using $($\ref{34}\,$)$ one can
find $a_1$ as a function of $a_2$:
 \be \label{a1}
a_1=\frac{-a_2\pm \sqrt{9g_2-3a_2^2\vphantom{I^{I^i}}}}{2} \,. \ee
Then with the aid of $($\ref{theorem}\,$)$ $d$ is expressed in
terms of $a_2$ also.
 Hence, the Bloch functions  $($\ref{45}\,$)$ and  $($\ref{46}\,$)$ are 
determined only by the
 eigenvalue $a_2$ which can take any value and,
 thus, these functions  generate the 1-parameter family of 2-SUSY
 invariant potentials mentioned in Corollary 1.
With appropriate modifications
the same is true for the functions $($\ref{47}$)$ and $($\ref{48}$)$.
\end{rem}

To illustrate  this  Remark we prove:
\begin{prop}
The potentials \be V_1(x)=V_0(x)+\Delta V(x) \,,\quad
V_0(x)=6\wp(x+\o ') \, ,\ee where
 \bea \label{1family}
 \Delta V(x) & = &
  -
 \displaystyle{\frac{[\wp(x+\o'+d)+
\wp(x+\o')]^2+2\wp(x+\o'+d)\wp(x+\o')-g_2/2}%
 {\wp(x+\o'+d)+\wp(x+\o')+\wp(d)}}
 \nonumber
 \\[1em]
& &
 +   3\wp(x+\o'+d)-3\wp(x+\o')-\wp(x-x_1)+\wp(x-x_2)\,,
 \eea
$d=\wp^{-1}(-\frac{a_1+a_2}{3})$, $x_1$, $x_2$ are defined as in
$($\ref{49}$)$, and $a_1$ in $($\ref{a1}$)$, form a one-parameter
family of real-valued two-gap potentials isospectral with $V_0$.
The role of the parameter is played by $a_2\le E_0$, where
$E_0=-\sqrt{3g_2}$ is the lowest band edge for  $V_0$.
\end{prop}

\noindent
{\it Proof}.
 The proof consists simply in calculating the right hand side of
  Eq. (\ref{21}) and taking its derivative which will give us
 the potential difference (see (\ref{1susy})).
 Using the expression (\ref{52}) below for $p$  in terms of the Weierstrass
  $\zeta$ function  and the addition formula (\ref{addition}) we first
 get
 \be
 \frac 12 (\log p)'=\frac{(\wp -\wt\wp )^2}{\wp ' +\wt\wp '}\, ,
 \ee
 where $\wp \equiv\wp (x+\o ')$ and
 $\wt\wp \equiv\wp (x+\o  '+d)$.
 To derive the final expression (\ref{1family}) we have used the
 formula (\ref{1op}) and the
 following properties of the Weierstrass functions:
 $(\wp ')^2=4\wp^3-g_2\wp-g_3$, $\wp ''=6\wp^2-g_2/2$
 and $\z '(x)=-\wp (x)$.
 The fact that these potentials have two-gaps follows from the
 fact that
 Darboux transformations preserve band structure (see e.g. \cite{Trl}).
 \hfill $\Box$

 It is interesting to observe that different representatives of
 this family look like displaced copies of the initial potential
 (see Fig.\ 1 $A$)  though $V_0$ is not 1-SUSY invariant which is clearly 
seen from Fig.\ 1 $B$, where the left hand side of  equation (\ref{11})
 is plotted.

\begin{center}
\epsfig{file=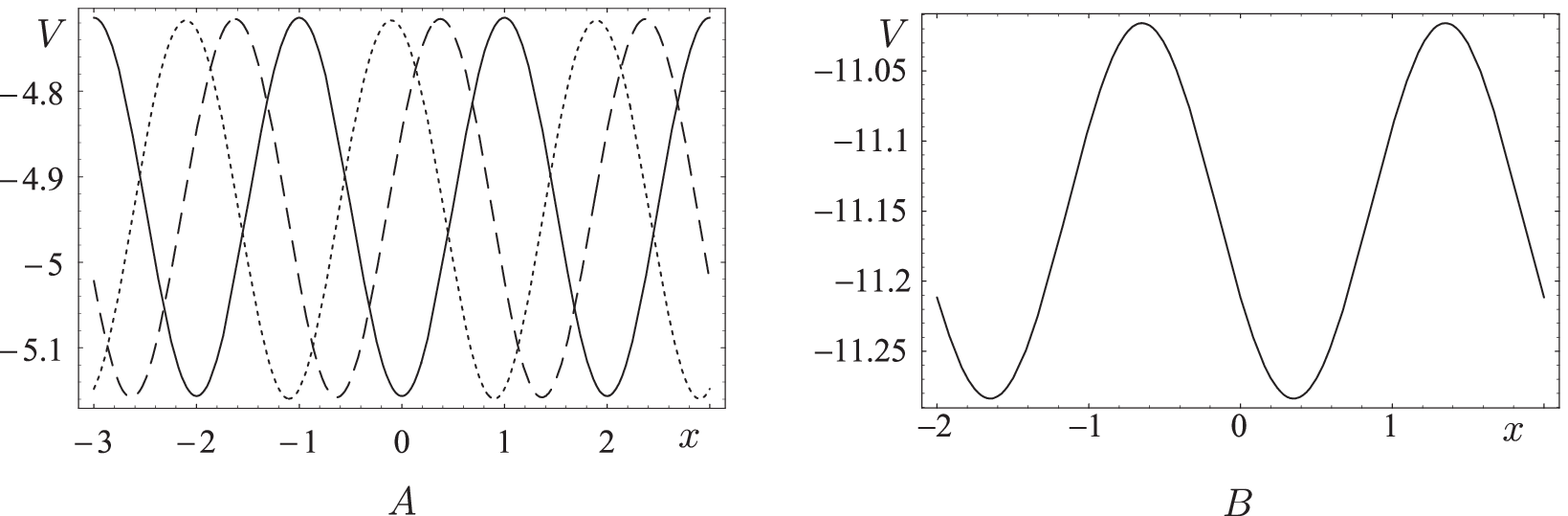}  
\label{fig1}
\medskip

\begin{minipage}{14cm}
{\small Figure 2:
$(A)$ Lam\'e-Ince potential
(solid line) at
 $\o =1$ and $\o '=2i$ and its SUSY partners.
 Dotted line corresponds to $a_2=-5$ and dashed line to $a_2=-6$;
 $(B)$ illustrates that the left hand side of the equation
 \protect{(\ref{11})} is not constant for $d=1.3$.
}
\end{minipage}
\end{center}
\medskip

\begin{center}
\epsfig{file=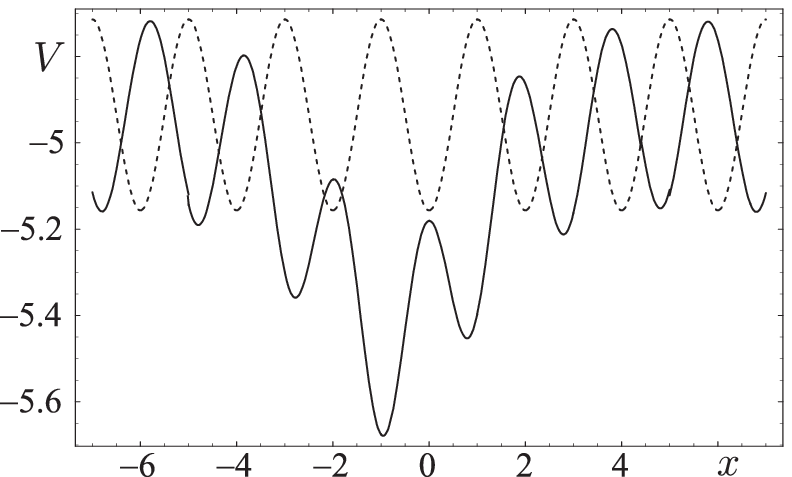}  
 \label{fig2}
\medskip

\begin{minipage}{14cm}
{\small Figure 3:
 Potential corresponding to the transformation function
 $u=u_2^-+0.5u_2^+$  at
 $\o =1$ and $\o '=2i$ and $a_2=-5.2$ (solid line). Dashed line
 shows the initial Lam\'e-Ince potential.
}
\end{minipage}
\end{center}
\medskip

Figure 3 shows a non-periodic potential obtained when a
 linear combination of the functions (\ref{45}) and (\ref{46}) is chosen
 as the transformation function. We stress that it possesses an energy
 level at $-5.2$ and the solution of the corresponding \Sc \ equation
 can easily be obtained by applying the first order Darboux
 transformation operator (\ref{1dt}) to the functions
 (\ref{45})--(\ref{48}).
 We would also like  to mention that the transformed
 potential tends to a displaced version of the initial potential
 at large values of $|x|$. This effect, first noticed in
 \cite{mielnik}, illustrates a nonlocal deformation produced by a
 Darboux transformation.

 Finally, we would like to point out that the function
 (\ref{45}) acquires a constant factor when its argument increases by
  the period $2\o $. This can be seen from
 properties of the Weierstrass $\sigma$ functions \cite{Bateman}
 that lead to
 \[
 u_2^-(x+2\o )=  u_2^-(x)
 \exp [\zeta (\o )(3d+x_2-x_1)-\o (3\zeta (d)-2b)]  \,.
 \]
 From here it follows, in particular, that for $a_2=3e_3$,
 $a_1=3e_2$, one has $d=\o $, $x_1=\o$ and $x_2=-2\o '$ which
 results in the condition
 $u_2^-(x+2\o )=u_2^-(x) \exp [2\zeta(\o ')\o-2\o '\zeta (\o  )]=
 -u_2^-(x)$, where we have used the Legendre relation \cite{Bateman}.
   This  result agrees with the fact that this
 function corresponds to the second band edge.

   In conclusion, we have characterized and investigated the
   interesting class of periodic potentials which are merely
   translated by a second order Darboux transformation. We have
   presented several examples, concentrating on the fascinating
   family
   of two-gap Lam\'e-Ince potentials. The nonlinear differential
   equation which must be satisfied by all such potentials, which
   may be related to the class of Painlev\'e equations, deserves
   further investigation.

\section*{Acknowledgments}

This work has been partially supported by the European FEDER and
by the Spanish MCYT (Grant BFM2002-03773), MECD (Grant
SAB2000-0240) and Junta de Castilla y Le\'on (VA085/02). MLG
thanks the Universidad de Valladolid for hospitality and support
and the NSF (USA) for partial support (Grant DMR-0121146). BFS
would like also to thank the Physics Department of CINVESTAV, where this
work was started in the spring of 2001, for kind hospitality.

\setcounter{equation}{0}
\renewcommand{\theequation}{A.\arabic{equation}}

\section*{Appendix A}

Here, to fix our notation, we relate the Weierstrass form of the
\Sc \ equation to its Lam\'e form (see also \cite{Bateman}, vol. 3).
According to  \cite{Bateman} one has
 \be
 \sn^2(z,k)=\frac{e_1-e_3}{\wp(x)-e_3}\,,\quad \wp(x+\omega
 ')=e_3+\frac{(e_3-e_2)(e_3-e_1)}{\wp (x)-e_3}
 \ee
 or
 \be \wp
 (x+\omega ')=e_3+(e_2-e_3)\sn^2(z,k)\, ,
 \ee
 where
 \be
k^2\equiv m=\frac{e_2-e_3}{e_1-e_3}\,,\quad z=(e_1-e_3)^{1/2}x\,.
\ee
Now, after replacing the variable $x$ by $z$ in the \Sc \ equation
with the potential (\ref{wp}), one gets
 \be
 \left[-
 \frac{d^2}{dz^2}+(6k^2\sn^2(z,k)-\wt E ) \right]\psi =0\, ,
 \ee
 where
 \be
 \wt E =\frac{E-6e_3}{e_1-e_3}=\frac{E}{e_1-e_3}+2m+2 \,.
 \ee
 Using this relation and Corollary 2 one recovers
  the known band edges for the Lam\'e equation (see e.g.\ \cite{fernandez}),
\bea
 &&\wt E_0= \frac{E_0-6e_3}{e_1-e_3}= 2m+2-2\delta ,\nonumber\\
&&\wt E_1=m+1, \qquad \wt E_{1'}=4m+1, \\
&&\wt E_{2}=m+4, \qquad \wt E_{2'}= 2m+2+2\delta  \,.\nonumber
\eea

\setcounter{equation}{0}
\renewcommand{\theequation}{B.\arabic{equation}}

\section*{Appendix B}

In this Appendix we  prove Proposition 3. In our notation the
addition theorem for the Weierstrass $\zeta$--function (see
\cite{Bateman}) \be \frac 12\, \frac{\wp '(x+d)+\wp '(x)}{\wp
(x+d)-\wp (x)}= \zeta(x)-\zeta(x+d)+\zeta(d) \ee becomes
\[
\frac 12\,\frac{S'}{D}=\zeta(x+\omega '+d)-\zeta(x+\omega ')-\zeta(d) \,.
\]
 Hence
 \be\label{52}
 p=3\frac{S'}{D}=6[\zeta(x+\omega '+d)-\zeta(x+\omega
 ')-\zeta(d)]\,.
\ee Recalling that \be
p^2=9\left(\frac{S'}{D}\right)^2=6S+36\wp(d) \ee
 we find
\be
S=6[\wp(x+\omega ')+\wp(x+\omega '+d)] \,.
\ee
Let us find now the expression for $1/p$.

The function $S$ has second order poles at the same points as
$\wp(x+\omega ')$ and $\wp(x+\omega '+d)$, that is at $x=-\omega
'$ and $x=-\omega '-d$. This means that $S$ is a second order elliptic
function and  has two second order zeros. Here and below, for
simplicity, we suppose that $d$ falls in the interval $(d_{\rm
min},d_{\rm max})$ indicated in Corollary 3.

Let us denote
\be f(x):=\wp(x+\omega ')+\wp(x+\omega '+d)+\wp(d)
\,.
\ee
 Then $p^2=36f\ge 0$ for real values of $\wp $.
 Note that for $d=\omega $, $f(0)=f(\omega )=e_1+e_2+e_3=0$. Hence, $x=0$
 and $x=\omega $ are the points of local minima for $f(x)$.
 Therefore $f'(0)=f'(\omega )=0$, i.e.
 the zeros are of  second order. Next, it is easy to see that
 $f'(\omega /2)=f'(3\omega /2)=0$ and the points $x=\omega /2$ and
 $x=3\omega /2$ are the points of local maxima for real valued $f(x)$.
 We find now the positions of these zeros.

Let $x_0$ be  a minimum (or a zero-point) for $f(x)$, that is
$f(x_0)=0$ and $f'(x_0)=0$. Let us abbreviate the notation by
putting
\[
\wp :=\wp(x_0+\omega ')\,,\quad \tilde \wp :=\wp(x_0+\omega '+d)\,,\quad
\wp_0:=\wp (d)  \,.
\]
Then $\wp +\tilde \wp +\wp_0=0$ and $\wp '+\tilde \wp '=0$ (with
the evident notation $\wp '=\wp '(x_0)$, \ldots). Or else
$\wp'^{\,2}-\tilde \wp '^{\,2}=0$. Using the differential equation
for $\wp $, $\wp'^{\,2}=4\wp^3-g_2\wp-g_3$, one finds from here
that
\[
(\wp -\tilde \wp)(4\wp ^2+4\tilde \wp ^2+4\wp\tilde \wp-g_2)=0  \,.
\]
Since $\wp \ne \tilde \wp $ this implies that
\[
0=4\wp^2+4\tilde\wp ^2+4\wp\tilde\wp-g_2=
4\wp^2+4\wp ^2_0+4\wp\wp_0-g_2
\]
and \be \wp (x_0+\omega ')=-\frac 12\wp _0\pm \frac
12\sqrt{g_2-3\wp_0^2} \,. \ee Hence, the positions $x_1$ and $x_2$
of the zeros of $f(x)$ are given by (\ref{49}). In using these
formulas it is  necessary to keep in mind that $\wp$ is a
multisheet function. Note, that for $d=\omega $ one gets $x_2=0$
and $x_1=\omega $ since
$\sqrt{g_2-3\wp_0^2\vphantom{i^i}}=e_2-e_3$, $\wp^{-1}(e_3)=\omega
'$, $\wp^{-1}(e_2)=\omega +\omega '$. Note also that since the
equation $\wp (z)=A$ defines $z$ up to a sign and up to the
periods, either the sum or the difference of $x_1$ and $x_2$
modulo periods is equal to $-d$.

The function $p\,(x)=\pm 6\sqrt{f(x)}$ has simple zeros  $x_1$ and
$x_2$ and simple poles $1/p\,(x)$. It is not difficult to find the
residues of $1/p\,(x)$ at these points \be
\begin{array}{lllll}
{\rm Res}&  \hspace{-.5em} {\displaystyle\frac{1}{p\,(x)}\,\,=}&
\hspace{-.5em} {-\,\rm Res}&\hspace{-.5em}
{\displaystyle\frac{1}{p\,(x)}\,\,=}&
{\displaystyle\frac{1}{\vphantom{I^{I^{I^I}}}6\sqrt{g_2-3\wp_0^2
\vphantom{o^{o^o}}}}}   \\[-1.5em] {{\scriptstyle x=x_1}}& &{\ \,
\scriptstyle x=x_2}& &
\end{array}   \,.
\ee This  implies that the residues of $({a_1-a_2})/{p}$ at the
points $x_{1,2}$ are $\pm 1/2$.

Now, using the known \cite{Bateman} decomposition of an elliptic
function in terms of $\zeta $ functions one finds \be \label{1op}
\frac{a_1-a_2}{p}=\frac 12\zeta(x-x_1)-\frac 12\zeta(x-x_2)+b  \,.
\ee The constant $b$ here may be calculated by the same formula at
$x=\omega '$ where $1/p=0$. This gives  (\ref{50}).

Using the fact that $\frac{\sigma '(z)}{\sigma (z)}=\zeta (z)$ one obtains
\be
\frac{a_1-a_2}{p}=
\frac 12\left[ \log \frac{\sigma (x-x_1)}{\sigma (x-x_2)} \right]'+b
\ee
and from (\ref{52}) one finds
\be
p=6\left[ \log \frac{\sigma (x+\omega '+d)}{\sigma (x+\omega ')} \right]'-
6\zeta (d) \,.
\ee

With the help of the last two relations one gets from (\ref{21})
\bea
{\displaystyle
[\log u_2^-(x)]'}\!\!=
{\displaystyle \frac 12[\log \,p\,]'+\frac 14\,p+\frac {a_1-a_2}{p}}
\\[1em]
= {\displaystyle \frac 12[\log \,p\,]'+
\frac 32\left[ \log \frac{\sigma (x+\omega '+d)}{\sigma (x+\omega ')} 
\right]'-
\frac 32\,\zeta (d)+
\frac 12\left[ \log \frac{\sigma (x-x_1)}{\sigma (x-x_2)} \right]'+b}\,.
\nonumber
 \eea
 It follows from here that (\ref{45}) is a quadrature of the
above formula. Similarly, the formulae (\ref{46})--(\ref{48})
follow from (\ref{45}) and (\ref{22})--(\ref{22b}), and the
proposition is proved.


\end{document}